# Sparse Structural Approach for Rating Transitions

Volodymyr Perederiy*

## Abstract

In banking practice, rating transition matrices have become the standard approach of deriving multi-year probabilities of default (PDs) from one-year PDs, the latter normally being available from Basel ratings. Rating transition matrices have gained in importance with the newly adopted IFRS 9 accounting standard. Here, the multi-year PDs can be used to calculate the so-called expected credit losses (ECL) over the entire lifetime of relevant credit assets.

A typical approach for estimating the rating transition matrices relies on calculating empirical rating migration counts and frequencies from rating history data. For small portfolios, however, this approach often leads to zero counts and high count volatility, which makes the estimations unreliable and unstable, and can also produce counter-intuitive prediction patterns such as non-parallel/crossing forward PD patterns.

This paper proposes a structural model which overcomes these problems. We make a plausible assumption of an underlying autoregressive mean-reverting ability-to-pay process. With only three parameters, this sparse process can well describe an entire typical rating transition matrix, provided the one-year PDs of the rating classes are specified (e.g. by the rating master scale).

The transition probabilities produced by the structural approach are well-behaved by design. The approach significantly reduces the statistical degrees of freedom of the estimated transition probabilities, which makes the rating transition matrix more reliable for small portfolios. The approach can be applied to data with as few as 50 observed rating transitions. Moreover, the approach can be efficiently applied to data consisting of continuous PDs (prior to rating discretization).

In the IFRS 9 context, the approach offers an additional merit: it can easily account for the macroeconomic adjustments, which are required by the IFRS 9 accounting standard.

---

* Perederiy Consulting (founder and consultant), PhD (Viadrina University, Germany)
research@perederiy-consulting.de, www.perederiy-consulting.de

# Acknowledgments

I would like to thank Jan-Philipp Hoffmann, PhD and Gerrit Reher, PhD for their inspirational contributions to this research.

# Introduction

Rating transition matrices are currently the standard tool for modeling long-term credit risk when only short-term rating data are available. A typical one-period rating transition matrix adopts the following form:

| From (initial) rating ($R_1$ ) | To rating ($R_2$ ) | | | | |
|---|---|---|---|---|---|
| | Rating 1 | Rating 2 | … | Rating $N_R$ | D |
| Rating 1 | $p_{1,1}$ | $p_{1,2}$ | … | $p_{1,N_R}$ | $p_{1,D} = PD_1$ |
| Rating 2 | $p_{2,1}$ | $p_{2,2}$ | … | $p_{2,N_R}$ | $p_{2,D} = PD_2$ |
| … | … | … | … | … | … |
| Rating $N_R$ | $p_{N_R,1}$ | $p_{N_R,2}$ | … | $p_{N_R,N_R}$ | $p_{N_R,D} = PD_{N_R}$ |
| D | 0 | 0 | 0 | 0 | 1 |

where $p_{R_1,R_2}$ stands for the probability of an obligor's transition to a rating $R_2$ given its initial rating $R_1$, over one period (normally one year), with $N_R$ non-default rating classes. Default is typically captured as a special rating class *D*. For the non-default rating classes, the transition probabilities to default are then the probabilities of default (PD) of the rating class. The transition probabilities from default to other rating classes are assumed to be zero (the default state thus being "absorbing"). Mathematically, the matrix is then square with the size $N_R + 1$, with all its elements being non-negative and with $PD_{R_1} + \sum_{R=1}^{N_R} p_{R_1,R} = 1$ holding for each row (initial rating) $R_1$.

The one-period matrices can then be extrapolated to multiple periods via simple matrix operations. In particular, given a one-year rating transition matrix $M$, the Y-year transition matrix $M_Y$ is easily calculated via raising it to the power of $Y$:

$$M_Y = M^Y = \prod_{m=1}^{Y} M \tag{1}$$

The Y-year cumulative PDs $CPD_{R_1,Y}$ for all initial rating classes $R_1$ (the probability of default occurring until the year $Y$) can then be extracted as the last column of the matrix $M_Y$ . From these cumulative PDs, other multi-year credit metrics can be easily calculated as follows:

- survival probabilities until the year $Y$: $SP_{R_1,Y} = 1 - CPD_{R_1,Y}$
- marginal (unconditional) PD for the year $Y$: $MPD_{R_1,Y} = CPD_{R_1,Y} - CPD_{R_1,Y-1}$
- forward (conditional) PD for the year $Y$: $FPD_{R_1,Y} = MPD_{R_1,Y}/SP_{R_1,Y-1}$

Such multi-year metrics are widely used for various risk management/controlling purposes, such as stress testing and risk-adjusted pricing. The importance of the multi-year metrics has recently increased due to the newly adopted IFRS 9 accounting standard, which requires the calculation of expected credit losses (ECL) over the entire lifetime of certain credit assets (see *IFRS Foundation [2014]*). Typically, the ECL for a relevant credit exposure is technically calculated as:

$$ECL_Y = \sum_{t=1}^{Y} EAD_t \;\; LGD_t \;\; MPD_t \;\; D_t \tag{2}$$

where $Y$ stands for the expected lifetime of the exposure (in years), $EAD_t$ the expected lifetime exposure at default, $LGD_t$ the expected lifetime loss given default, $D_t$ the discount factor, and $MPD_t$ the marginal probability of default for the year $t$. Thus, the ECL calculation requires the knowledge of multi-year marginal PDs which can be derived from the matrices as described above[1].

The one-year transition matrices *M* are typically estimated as empirical frequencies of annual rating transitions observed over a few years for all obligors in a credit portfolio (also known as 'cohort' method, see e.g. *Gerhold et al [2017]*). Another popular and slightly more advanced approach lies in attempting to estimate the continuous-time (instantaneous) rating transition rates (also known as 'generator' matrices) and deriving the one-year transition probabilities from these rates (see *Lando and Skødeberg [2002]*).

However, for small portfolios, with only a few dozen rating transitions observed, because of the underling probabilistic nature of transitions, such empirical estimators are noisy, volatile, and often show gaps (zero frequencies) and monotony reversions in the frequencies[2]. A typical remedy is to then smooth the empirical transition frequencies with one of numerous techniques (e.g. splines, normal quantiles etc.) and to inter/extrapolate them into rating classes where empirical data are absent.

Rather than such technical smoothing, this paper proposes a structural approach to deal with these problems related to small portfolios. The whole empirical rating transition matrix is thereby reduced to an underlying process of the 'ability-to-pay', with very few parameters needed. This process is stochastic and autoregressive with a mean reversion, which makes it a good approximation of reality. As only a few parameters are needed, the structural transition estimators are considerably sparser than the direct empirical transition estimators.

The observed empirical rating transitions can then be seen as realizations of this stochastic process and thus can be used to estimate its parameters using MLE (maximum likelihood estimation). Once the corresponding process parameters are estimated, the regularized one-year transition matrices can be easily derived from the process (as transition probabilities rather than observed frequencies).

# Model

## General Model Specification

We start with an autoregressive specification for the ability-to-pay (AP) process of an obligor:

$$\begin{gathered} AP_{t+1} = a_0 + a_1 AP_t \; + r_{t+1} \\ r_t \sim F \end{gathered} \tag{3}$$

With random, identically and independently distributed returns $r_t$ characterized by cumulative distribution function (cdf) $F$ and expected value $E(r_t) = 0$. This is equivalent to an assumption that the expected AP change is linearly dependent on the current AP level:

$$E(AP_{t+1} - AP_t) = a_0 + (a_1 - 1)AP_t \; = (1 - a_1)\left(\frac{a_0}{1 - a_1} - AP_t\right) \tag{4}$$

and thus the AP is reverting to its mean $\frac{a_0}{1-a_1}$, with intensity of mean reversion determined by $1 - a_1$. For $0 \leq a_1 < 1$, the ability-to-pay process is stationary.

The obligor is assumed to default if its ability-to-pay falls below 0. In this context, the ability-to-pay can be interpreted as a logarithm of the ratio of assets and liabilities of the obligor; the zero threshold thus corresponding to liabilities exceeding assets. Thus, the probability of default at a time point $t$ can be calculated as follows:

$$PD_t = P(AP_{t+1} < 0) = P(r_{t+1} < -a_0 - a_1 AP_t) = F(-a_0 - a_1 AP_t) \qquad (5)$$

[1] Furthermore, the multi-year forward PDs are often used in the IFRS 9 context as a comparison criterion for identifying a significant deterioration of credit quality, which triggers the application of the multi-year ECL.

[22] More generally, from a statistical point of view, both methods (cohort and generator) suffer from the 'overfitting' problem, as the number of available data (transition counts) is low in comparison to the number of estimated parameters (transition probabilities).

The above mean-reverting specification is a realistic assumption for the asset-liability ratio of an obligor: extremely high ratios are quite unlikely to persist, because they impair the advantages of tax deduction and profitability leverage for businesses. Extremely low ratios are also disadvantageous, as business conduction becomes difficult for companies with very high credit risk. Thus, a certain long-term mean/equilibrium level for the ability-to-pay can be assumed to exist; this level is likely to depend on the obligor`s industry, legal form and (possibly) size.

From (5), the implied level of the ability-to-pay can be inferred from a known PD as follows:

$$AP_t = -\frac{1}{a_1}\left(\mathrm{F}^{-1}(PD_t) + a_0\right) \qquad (6)$$

Now, for the PD of the next time point $t+1$, it holds:

$$\begin{gathered} PD_{t+1} = F(-a_0 - a_1 AP_{t+1}) = \\ F\left(-a_0 - a_1\left(a_0 + a_1 AP_t + r_{t+1}\right)\right) = \\ F\left(-a_0 - a_1\left(a_0 + a_1\left(-\frac{1}{a_1}\left(\mathrm{F}^{-1}(PD_t) + a_0\right)\right) + r_{t+1}\right)\right) = \\ F\left(-a_0 - a_1\left[-\mathrm{F}^{-1}(PD_t) + r_{t+1}\right]\right) \end{gathered} \qquad (7)$$

As shown in Appendix, the conditional distribution of the (future) $PD_{t+1}$, given the knowledge of the (current) $PD_t = \widehat{PD}_t$, can be derived analytically. The pdf (probability density function) is:

$$pdf(PD_{t+1}) = \frac{1}{a_1}\frac{f\left(\mathrm{F}^{-1}(\widehat{PD}_t) - \frac{\mathrm{F}^{-1}(PD_{t+1}) + a_0}{a_1}\right)}{f\left(\mathrm{F}^{-1}(PD_{t+1})\right)} \qquad (8)$$

where $f$ stands for the pdf of the return $r_t$, and the hat operator ( $\widehat{\ }$ ) denotes values (realizations) of random variables. The cdf (cumulative distribution function) is:

$$cdf(PD_{t+1}) = 1 - F\left(\mathrm{F}^{-1}(\widehat{PD}_t) - \frac{\mathrm{F}^{-1}(PD_{t+1}) + a_0}{a_1}\right) \qquad (9)$$

or alternatively (for return distributions with symmetric densities):

$$cdf(PD_{t+1}) = F\left(\frac{\mathrm{F}^{-1}(PD_{t+1}) + a_0}{a_1} - \mathrm{F}^{-1}(\widehat{PD}_t)\right) \qquad (10)$$

The probability of default happening between $t$ and $t+1$ is $PD_t$, as previously defined. We encode actual default events by setting $\widehat{PD}_{t+1} = 100\%$. We also make the classical assumption that default is an absorbing state, i.e. once defaulted, the obligor remains in this status during the following periods. This is equivalent to overriding the above process specification with a special case:

$$P(PD_{t+1} = 100\% \mid PD_t = 100\%) = 1 \qquad (11)$$

It should be noted that the presented AP process specification imposes a natural upper limit for the PD of a non-defaulted obligor. This maximum probability of default $PD_{max}$ is achieved, for a time point $t$, if $AP_t = 0$, which, when combined with (5), amounts to:

$$PD_{max} = F(-a_0) \qquad (12)$$

It is important to note that the above model specification is essentially invariant with respect to a non-zero ability-to-pay threshold and a scaling of the return[3]. For this reason, the threshold and scaling do not offer additional fitting advantages in this model specification; they need not (and also can not) be separately estimated from data.

## Distribution of Return

The choice of the return distribution cumulative/cdf function $F$ (or alternatively density/pdf function $f$) is an important issue. The normal distribution proves to be unsuitable, leading to unrealistic transition probabilities in this model framework. Fat tail distributions can produce plausible rating transition probabilities (see *Löffler [2002]*). In this paper, we use the t distribution with a single degrees-of-freedom parameter $df$. It is described by the following pdf and cdf distribution functions:

$$pdf(t) = \frac{\Gamma\left(\frac{df+1}{2}\right)}{\sqrt{\pi\, df}\,\Gamma\left(\frac{df}{2}\right)}\left(1+\frac{t^2}{df}\right)^{-\frac{df+1}{2}}$$

$$cdf(t) = \frac{1}{2} + t\,\Gamma\left(\frac{df+1}{2}\right)\frac{H\left(\frac{1}{2},\frac{df+1}{2};\frac{3}{2};-\frac{t^2}{df}\right)}{\sqrt{\pi\, df}\,\Gamma\left(\frac{df}{2}\right)} \tag{13}$$

where $\Gamma$ stands for the gamma function, $H$ for the hypergeometric function, and $df$ for the degrees-of-freedom parameter.

The t distribution has the advantage that the tail fatness can be controlled by the parameter $df$, which can be continuous. The mean (expected value) exists for $df > 1$ and equals 0. The variance exists for $df > 2$ and equals $df/(df-2)$. The distribution is symmetric.

Additionally, as $df \to \infty$, the t-distribution converges to the (standard) normal distribution. Thus, normal return distribution can be regarded as a special case for this model specification.

# Maximum Likelihood Estimation

Given the analytical expressions (8) and (9), the method of maximum likelihood estimation (MLE) can be applied to estimate the parameters. The exact estimation approach depends on the data available. In particular, we will distinguish between the cases when undiscretized (continuous) PDs vs. discretized (rating) PDs are available as data. We will also distinguish the cases when default events are not included in the data (living portfolios).

## Transitions between Continuous PDs

In many cases, the continuous PDs of obligors would be observable. This would, for example, be the case for probabilistic (e.g. logit/probit) rating-model outputs prior to discretization. In this case, the data could consist of e.g. $N$ observations of yearly transitions from $\widehat{PD}_{i,t}$ to $\widehat{PD}_{i,t+1}$ for $i = 1, \dots, N$, with $i$ representing obligors and (possibly) years, and defaults encoded as $\widehat{PD}_{i,t+1} = 100\%$.

The likelihood of each transition then corresponds to the pdf value:

$$L_i = \frac{1}{a_1}\frac{f\left(\mathrm{F}^{-1}\left(\widehat{PD}_{i,t}\right) - \frac{\mathrm{F}^{-1}\left(\widehat{PD}_{i,t+1}\right) + a_0}{a_1}\right)}{f\left(\mathrm{F}^{-1}\left(\widehat{PD}_{i,t+1}\right)\right)} \tag{14}$$

---

[3] In particular, for a threshold $T$ and scaling factor $S$, the PD is defined as $PD_t = P(AP_{t+1} < T)$ with $AP_t = a_0 + a_1 AP_{t-1} + S r_t$. The formulae presented remain valid, except that the term $a_0$ is replaced by the term $(a_0 - T + a_1 T)/S$.

with the special case:

$$L_i = \widehat{PD}_{i,t} \;\; if \;\; \widehat{PD}_{i,t+1} = 100\% \tag{15}$$

The MLE then corresponds to the maximization of the product of the likelihoods across all transitions:

$$L = \prod_{i=1}^{N} L_i \tag{16}$$

with respect to the autoregressive parameters $a_0$, $a_1$ and the parameters of the distribution $F$ (a single $df$ parameter for the t distribution). Normally, the sum of the logarithms of the likelihoods (log-likelihood) is technically maximized instead of the product:

$$\begin{gathered} a_0, a_1, df = \underset{a_0, a_1, df}{\arg\max} \sum_{i=1}^{N} LL_i \\ LL_i = \ln f\left(\mathrm{F}^{-1}\left(\widehat{PD}_{i,t}\right) - \frac{\mathrm{F}^{-1}\left(\widehat{PD}_{i,t+1}\right) \;\; + a_0}{a_1}\right) - \ln f\left(\mathrm{F}^{-1}\left(\widehat{PD}_{t+1}\right)\right) - ln a_1 \\ LL_i = ln\widehat{PD}_{i,t} \;\; if \;\; \widehat{PD}_{i,t+1} = 100\% \end{gathered} \tag{17}$$

If defaults are not included in the observable transition data (living portfolio), the special case (15) is not relevant, and the likelihood generally corresponds to the pdf (14) divided by the survival probability $(1 - \widehat{PD}_{i,t})$ which results, after simplifications, in the same expression as in (17) for the sum of logarithms to be maximized.

Note that, given $\widehat{PD}_{i,t+1}$ and the three estimated parameters, the likelihood function (14) is unimodal (∩-shaped) in $\widehat{PD}_{i,t}$, with the likelihood maximum achieved at:

$$PD_{i,t}^{mode} = F\left(\frac{\mathrm{F}^{-1}\left(\widehat{PD}_{i,t+1}\right) \;\; + a_0}{a_1}\right) \tag{18}$$

Also, given $\widehat{PD}_{i,t}$ and the three estimated parameters, the likelihood function (14) is unimodal in $\widehat{PD}_{i,t+1}$, with the likelihood maximum achieved at:

$$PD_{i,t+1}^{mode} \approx F\left(-a_0 + a_1 \mathrm{F}^{-1}\left(\widehat{PD}_{i,t}\right)\right) \tag{19}$$

with very close approximation. These properties will prove later to be beneficial for the model output.

## Transitions from Continuous PDs to PD Intervals

In this case, the transitions observed are from a continuous $\widehat{PD}_{i,t}$ in $t$ to a PD interval $(\widehat{PD}_{i,t+1}^{low}, \widehat{PD}_{i,t+1}^{high})$ in $t+1$. The likelihood of observing each such transition can be easily obtained as the difference of corresponding cdfs:

$$\begin{gathered} L_i = P\left(\widehat{PD}_{i,t+1}^{low} < PD_{i,t+1} < \widehat{PD}_{i,t+1}^{high}\right) = \\ P\left(PD_{t+1} < \widehat{PD}_{i,t+1}^{high}\right) - P\left(PD_{t+1} < \widehat{PD}_{i,t+1}^{low}\right) = \\ = F\left(\mathrm{F}^{-1}\left(\widehat{PD}_{i,t}\right) - \frac{\mathrm{F}^{-1}\left(\widehat{PD}_{i,t+1}^{low}\right) + a_0}{a_1}\right) - F\left(\mathrm{F}^{-1}\left(\widehat{PD}_{i,t}\right) - \frac{\mathrm{F}^{-1}\left(\widehat{PD}_{i,t+1}^{high}\right) + a_0}{a_1}\right) \end{gathered} \tag{20}$$

with the special case:

$$L_i = \widehat{PD}_{i,t} \;\; if \;\; \widehat{PD}_{i,t+1} = 100\% \tag{21}$$

Maximum likelihood is then equivalent to maximizing the product of these transition probabilities across all transitions observed, or alternatively, equivalent to maximizing the sum of the logarithms of these probabilities, with respect to the parameters $a_0$, $a_1$ and the parameters of the distribution $F$ (a single $df$ parameter for the t distribution). Again, the expression to be maximized remains unchanged if defaults are not included in the observable transition data (living portfolio).

Two further special cases in this context are the conditions $\widehat{PD}^{low}_{i,t+1} = 0$ and $\widehat{PD}^{high}_{i,t+1} = PD_{max} = F(-a_0)$.

With $\widehat{PD}^{low}_{i,t+1} = 0$, the expression (20) reduces to:

$$\begin{gathered} P\left(\widehat{PD}^{low}_{i,t+1} < PD_{i,t+1} < \widehat{PD}^{high}_{i,t+1}\right) = \\ P\left(PD_{t+1} < \widehat{PD}^{high}_{i,t+1}\right) = \\ = 1 - F\left(\mathrm{F}^{-1}\left(\widehat{PD}_{i,t}\right) - \frac{\mathrm{F}^{-1}\left(\widehat{PD}^{high}_{i,t+1}\right) + a_0}{a_1}\right) \end{gathered} \tag{22}$$

From here, by setting $\widehat{PD}^{high}_{i,t+1} \equiv \widehat{PD}_{i,t}$, we can obtain the probability of a PD decrease as:

$$P(PD_{decr}) = P\left(\widehat{PD}_{i,t+1} < \widehat{PD}_{i,t}\right) = 1 - F\left(\mathrm{F}^{-1}\left(\widehat{PD}_{i,t}\right) - \frac{\mathrm{F}^{-1}\left(\widehat{PD}_{i,t}\right) + a_0}{a_1}\right) \tag{23}$$

In particular, for symmetric distributions, with $F(0) = 0.5$, the PD-decrease probability equal to 50% is reached at the equilibrium point:

$$\widehat{PD}_{i,t} = PD_{eq} = F\left(\frac{a_0}{a_1 - 1}\right) \tag{24}$$

From this, the mean-reverting pattern also becomes evident for the PD. As $\widehat{PD}_{i,t}$ increases, the probability of a PD-decrease becomes higher (for $0 < a_1 < 1$), reaching 50% at the equilibrium point.

With the second special case $\widehat{PD}^{high}_{i,t+1} = PD_{max}$, the expression (20) reduces to:

$$\begin{gathered} P\left(\widehat{PD}^{low}_{i,t+1} < PD_{i,t+1} < \widehat{PD}^{high}_{i,t+1}\right) = \\ = F\left(\mathrm{F}^{-1}\left(\widehat{PD}_{i,t}\right) - \frac{\mathrm{F}^{-1}\left(\widehat{PD}^{low}_{i,t+1}\right) + a_0}{a_1}\right) - \widehat{PD}_{i,t} \end{gathered} \tag{25}$$

From here, by setting $\widehat{PD}^{low}_{i,t+1} \equiv \widehat{PD}_{i,t}$, we can obtain the probability of a PD-increase (up to the non-default PD maximum) as:

$$P(PD_{incr}) = F\left(\mathrm{F}^{-1}\left(\widehat{PD}_{i,t}\right) - \frac{\mathrm{F}^{-1}\left(\widehat{PD}_{i,t}\right) + a_0}{a_1}\right) - \widehat{PD}_{i,t} \tag{26}$$

## Transitions from Discrete PDs to Discrete PDs (Ratings)

The third type of data available consists of transitions between discretized PDs for each obligor $i$, i.e. from an interval $(\widehat{PD}^{low}_{i,t}, \widehat{PD}^{high}_{i,t})$ to an interval$(\widehat{PD}^{low}_{i,t+1}, \widehat{PD}^{high}_{i,t+1})$. An exact calculation of the transition probability is difficult in this case. Firstly, the distribution of the (unknown continuous) $PD_{i,t}$ over interval $(\widehat{PD}^{low}_{i,t}, \widehat{PD}^{high}_{i,t})$ would have to be accounted for. Secondly, this distribution would also depend on the previously observed PD intervals of the obligor $(\widehat{PD}^{low}_{i,t-1}, \widehat{PD}^{high}_{i,t-1})$, $\left(\widehat{PD}^{low}_{i,t-2}, \widehat{PD}^{high}_{i,t-2}\right)$, …, so that transitions are, strictly speaking, non-Markovian here.

For most practical purposes however, simple approximations can be applied. In particular, for sufficiently narrow PD intervals, the assumption $\widehat{PD}_{i,t} = \frac{1}{2}\widehat{PD}^{low}_{i,t} + \frac{1}{2}\widehat{PD}^{high}_{i,t}$ can be made and (20) can be applied correspondingly.

The transition data available in the context of Basel rating systems may show some special characteristics. The rating systems developed within Basel IRB regulations (internal rating-based models for capital adequacy of banks) are typically built via PD discretization. In particular, the PD space is divided into $N_R$ intervals (typically 10-20) according to the rating 'master scale'. Each rating $R$ is then defined by the corresponding PD interval $(pd_R^{low}, pd_R^{high})$, with rating $R$ assigned to the obligor if the continuous rating-model PD output falls between $pd_R^{low}$ and $pd_R^{high}$. Also, an additional fixed "assigned" PD $pd_R^{asn}$, with $pd_R^{low} < pd_R^{asn} < pd_R^{high}$, is often defined in the master scale for each rating class (and used for the actual calculation of the Basel capital requirements).

The transition data consists then often of an empirical rating transition matrix. With such data, there are observed cumulated counts $N_{R_1,R_2}$ for transitions from an initial rating $R = R_1$ (in a period $t$) to a rating $R = R_2$ (in the next period $t + 1$) of various obligors. We use $P_{R_1,R_2}$ for the corresponding conditional transition probability that an obligor with an initial rating $R_1$ receives the rating $R_2$ in the next period. Then, the likelihood of the observed counts $N_{R_1,R_2}$ can be calculated using the multinomial distribution. In particular, for each initial rating $R_1$, the probability of the observed counts is:

$$\mathrm{P}(R_1) = N_{R_1}! \prod_{R_2=1}^{N_R} \frac{P_{R_1,R_2}^{N_{R_1,R_2}}}{N_{R_1,R_2}!}$$
$$N_{R_1} \equiv \sum_{R_2=1}^{N_R} N_{R_1,R_2} \tag{27}$$

Then, the overall log-likelihood of the observed counts (over all initial ratings) can be calculated as follows:

$$LL = ln \prod\nolimits_{R_1=1}^{N_R} \mathrm{P}(R_1) =$$
$$\sum\nolimits_{R_1=1}^{N_R} lnN_{R_1}! + \sum\nolimits_{R_1=1,R_2=1}^{N_R} N_{R_1,R_2} lnP_{R_1,R_2} - \sum\nolimits_{R_1=1,R_2=1}^{N_R} lnN_{R_1,R_2}! \tag{28}$$

As the first and the third terms do not depend on $P_{R_1,R_2}$, maximizing the log-likelihood is equivalent to maximizing the term $\sum N_{R_1,R_2} lnP_{R_1,R_2}$. The transition probabilities $P_{R_1,R_2}$ can be specified using the above conclusions for the transitions from a continuous PD to a PD interval. In particular, the assigned rating PD can be used as the initial PD, which results in:

$$P_{R_1,R_2} =$$
$$F\left(\mathrm{F}^{-1}\left(pd_{R_1}^{asn}\right) - \frac{\mathrm{F}^{-1}\left(pd_{R_2}^{low}\right) + a_0}{a_1}\right) -$$
$$F\left(\mathrm{F}^{-1}\left(pd_{R_1}^{asn}\right) - \frac{\mathrm{F}^{-1}\left(pd_{R_2}^{high}\right) + a_0}{a_1}\right) \tag{29}$$

If the assigned PDs $pd_{R_1}^{asn}$ are not available, mid values $pd_{R_1}^{mid} = \frac{1}{2}\left(pd_{R_1}^{low} + pd_{R_1}^{high}\right)$ can be instead applied. The two special cases are transitions to the best rating class with $pd_{R_2}^{low} = 0$, and transitions to the worst rating class with $pd_{R_2}^{high} = PD_{max}$. Here, (22) and (25) can be applied accordingly.

Thus, the log-likelihood can be maximized, using an empirical rating transition matrix, with respect to the parameters $a_0$ and $a_1$, as well as the parameters of the return distribution $F$.

# Practical Issues and Usage

## Typical Parameter Estimates

We estimated the proposed model on a number of real rating transition datasets (with between approx. 50 and 1000 rating transitions included). Technically, the SAS procedure *NLMIXED* was used for the likelihood maximization. The estimations delivered plausible results. In particular, the autoregressive coefficient $a_1$ took realistic values of 0.7 to 0.95, confirming the expectation of a mean reverting character of obligor creditworthiness. The degrees-of-freedom estimator $df$ of the t distribution was between 2 and 5, clearly indicating fat tails in the implied return distribution. The theoretically derived equilibrium PD (see (24)) was close to the average portfolio PDs. The MLE convergence was fast and unproblematic for the portfolios inspected.

## Regularized One-Year Transition Matrices

Once the process parameters $\hat{a}_0$, $\hat{a}_1$ and $\widehat{df}$ have been estimated, the process specification in (3)-(5) can be exploited to draw on the distribution of future PDs based on a known current PD. For one-year predictions $PD_{i,t+1}$, this distribution is straightforward and equivalent to the likelihood function used for the parameter estimation equation. In particular, for rating data, the regularized one-year probability of an obligor belonging in one year to the rating $R_2$ given its current rating $R_1$ can be readily calculated as in (29):

$$P_{R_1,R_2}^{reg} = F_{\widehat{df}}\left(F_{\widehat{df}}^{-1}\left(pd_{R_1}^{asn}\right) - \frac{F_{\widehat{df}}^{-1}\left(pd_{R_2}^{low}\right) + \hat{a}_0}{\hat{a}_1}\right) - F_{\widehat{df}}\left(F_{\widehat{df}}^{-1}\left(pd_{R_1}^{asn}\right) - \frac{F_{\widehat{df}}^{-1}\left(pd_{R_2}^{high}\right) + \hat{a}_0}{\hat{a}_1}\right) \tag{30}$$

with, analogously, special cases for the best and worst rating classes.

Unlike the observed transition counts $N_{R_1,R_2}$ or observed transition frequencies $N_{R_1,R_2}/N_{R_1}$, the regularized transition probabilities $P_{R_1,R_2}^{reg}$ offer some important advantages: structural soundness, completeness (positive/non-zero transition probabilities), and monotonically decaying probabilities (horizontally and vertically). The latter property results from the properties (18) and (19) in the continuous case; these properties also generally remain valid after discretization into typical rating intervals.

## Multiyear Forecasts

For multi-year predictions (with a prediction horizon $Y > 1$), a closed-form expression for the distribution of a future probability of default $PD_{i,t+Y}$ does not exist. In principle, it could be derived directly from the process (3) via determining the distribution of the future ability-to-pay. In particular, the future ability to pay $AP_{i,t+Y}$ can be deduced from the current ability-to-pay $AP_{i,t}$ and the subsequent returns as:

$$AP_{i,t+Y} = AP_{i,t}\, a_1{}^{Y} + \frac{a_0\left(1 - a_1{}^{Y}\right)}{1 - a_1} + \sum_{k=1}^{Y} a_1{}^{k-Y} r_{t+k} \tag{31}$$

Thus, it depends on the weighted sum of returns between $t + 1$ and $t + Y$. With the t distribution being assumed for each return, the analytical distribution of this sum is, however, unknown; it can only be estimated via simulation. Moreover, due to the assumed absorbing nature of defaults, the obligor would be defaulted by the time $t + Y$ (i.e. $PD_{i,t+Y} = 100\%$) with the probability of $\min\left(AP_{i,t+1}, \dots, AP_{i,t+Y}\right)$ falling below 0, which also cannot be derived analytically.

Due to these complications, the usual procedure of deriving the multi-year credit metrics via matrix powers/exponentiation of the one-year transition matrix (as outlined in the introduction) can also be recommended for the structural approach. However, the one-year matrix used here should be the regularized matrix derived from the estimated parameters as shown above in (30). Note that the transition matrix in (30) might have an arbitrary granularity of rating classes, possibly different from that of the rating transition data used for the parameter estimation in (29). In fact, the finer the granularity in (30), the more exact the approximation of the multi-year dynamics of the continuous process (31) will be.

Finally, some important properties are obvious from (31). In particular, for $a_1 > 0$, the higher the current $AP_{i,t}$ is, the higher, on average, the future $AP_{i,t+Y}$ would be. Due to the monotonous relationship $PD = F(-a_0 - a_1 AP)$, this property also extends to the PDs. This leads to an important property of non-intersecting forward PDs (and, as a consequence, non-intersecting cumulative PDs) produced by the process: the higher the initial PD of an obligor, the higher its forward PDs.

## Macroeconomic Adjustments

Most rating models are hybrid, combining through-the-cycle and point-in-time features. IFRS 9 generally requires considering the current and future macroeconomic conditions when using the lifetime PDs.

The autoregressive parameter $a_0$ in the process specification is the natural choice for reflecting the macroeconomic situation. Via (23) and (26), it directly influences the rating upgrade (PD decrease) and rating downgrade (PD increase) rate which is often seen as the measure of macroeconomic conditions when applied to rating systems. With several years of hybrid-rating transition data, the estimated parameter $a_0$ would roughly correspond to the long-term average of macroeconomic conditions. In order to account for a specific macroeconomic development, this parameter may need to be adjusted. This can be done e.g. via factoring in the upgrade/downgrade rate which is expected under this macroeconomic situation, using the relationships in (23) and (26).

Another possibility to account for the macroeconomic situation is to split the return $r_{i,t}$ of each obligor *i* as a sum of a systematic (macroeconomic) return $s_t$ and an idiosyncratic (obligor-specific) return $\varepsilon_{i,t}$. Classical asset-based credit portfolio models (see e.g. *Vasicek [2002]*) use this approach, assuming normal return distributions. With the t distribution, however, the split is considerably more difficult, as the sum of t distributions is a non-standard distribution (and thus, is not covered by standard statistical packages) and does not have an analytical representation.

# Performance and Value Added: Simulation Study

The ultimate proof of the advantages of the presented approach in the IFRS 9 context lies in an improved prediction of long-term (lifetime) PDs for small portfolios, in comparison to using the naïve approach of empirical transition matrices.

With this purpose in mind, we conducted the following simulation study. Firstly, a large realistic credit portfolio of obligors with a median PD of approx. 0.5% and a lognormal PD distribution was generated, and initial ratings were assigned to these obligors based on their PDs, using a logarithmically-built rating master scale of 20 ratings. Then, the transitions of the PDs and ratings over a 10-year period, along with defaults, were independently simulated for each obligor in the portfolio, using the process specifications (3) and (5) with parameters $a_0 = 1.2, a_1 = 0.8$ and $df = 3.5$. Using the simulated defaults, the cumulative 10-year default rate was calculated for each initial rating class and served as its “true” 10-year cumulative PD.

Then, the predictions of the structural vs. empirical approaches were compared, for a large portfolio of one million observed rating transitions. In particular, the empirical rating transition (frequency) matrix was first constructed as usual using the cohort method[4]. Then, based on this empirical matrix, the process (3) was fitted using the MLE estimation as described in (29). Then, the structural/regularized one-year transition matrix was calculated from the estimated parameters as in

[4] The one-year frequency of transitions to the default state was assumed to be known and equal to the assigned PD of the rating class.

(30). Then, from the two one-year transition matrices (empirical and regularized), through the matrix operations described in the introduction, the forward PDs were calculated for the prediction horizons of up to 10 years, for each of the 20 initial rating classes. The following figures depict these forward probabilities produced by the structural/regularized vs. empirical/naïve matrices for this large portfolio.

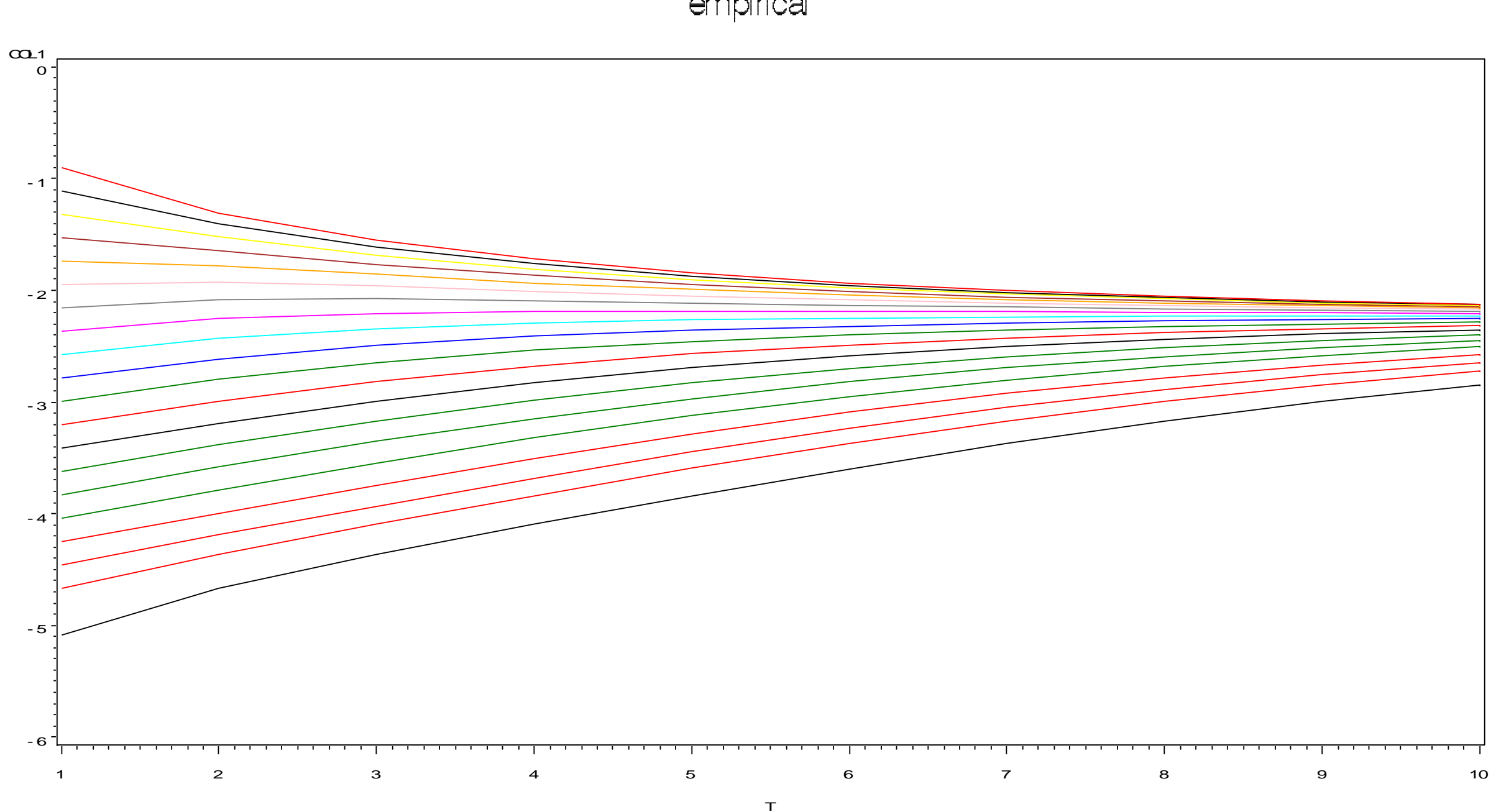

*Predictions of the forward one-year default probabilities produced by the empirical vs. structural approaches. X-axis: prediction horizon (years), Y-axis: forward PD (log10-scaled) produced by the empirical transition matrix for the 20 initial rating classes.*

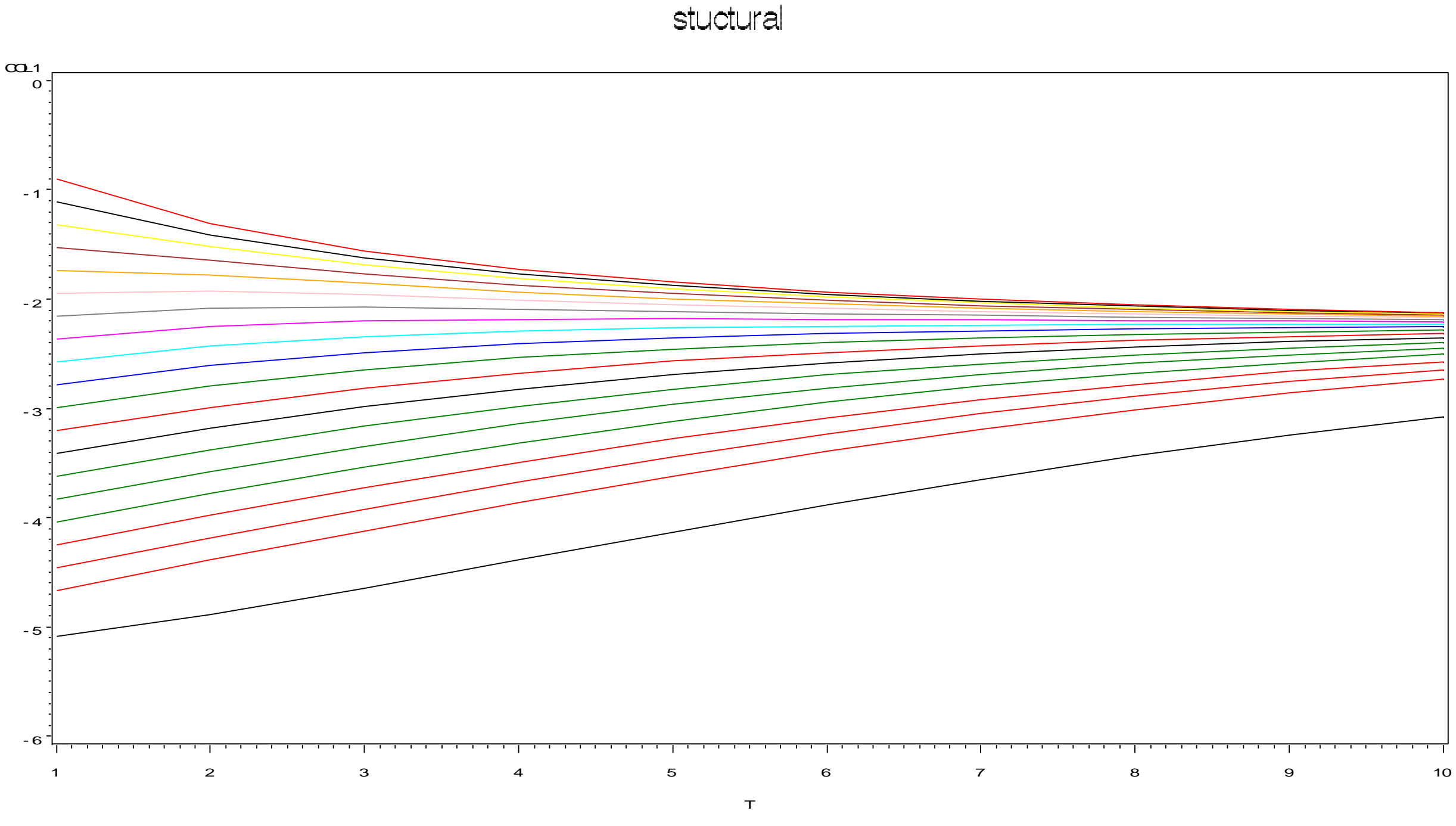

*Predictions of the forward one-year default probabilities produced by the empirical vs. structural approaches. X-axis: prediction horizon (years), Y-axis lower figure: forward PD (log10-scaled) produced by the structural transition matrix for the 20 initial rating classes.*

It is clear that, for this large portfolio, these two approaches generally produce very similar multi-year PDs. A slight difference is only seen for the best (lowest-PD) rating class.

Now, the performance of the empirical vs. structural approaches was compared for small portfolios. To achieve this, 100 small samples, each consisting of only 100 one-year rating transitions, were randomly drawn from the above large population. Each sample should represent rating transitions which are observed in a small portfolio. For each sample, the empirical/naïve one-year rating transition (frequency) matrix and structural/regularized one-year transition matrix were first constructed analogously to the large portfolio[5]. Again, the multi-year metrics were then derived from the two transition matrices for each sample.

As expected, the empirical vs. structural approaches showed a significantly different performance for the small portfolios. The following graph compares the "true" 10-year cumulative PD with its empirical vs. structural predictors, aggregating all 100 samples drawn, for each of the 20 initial rating classes.

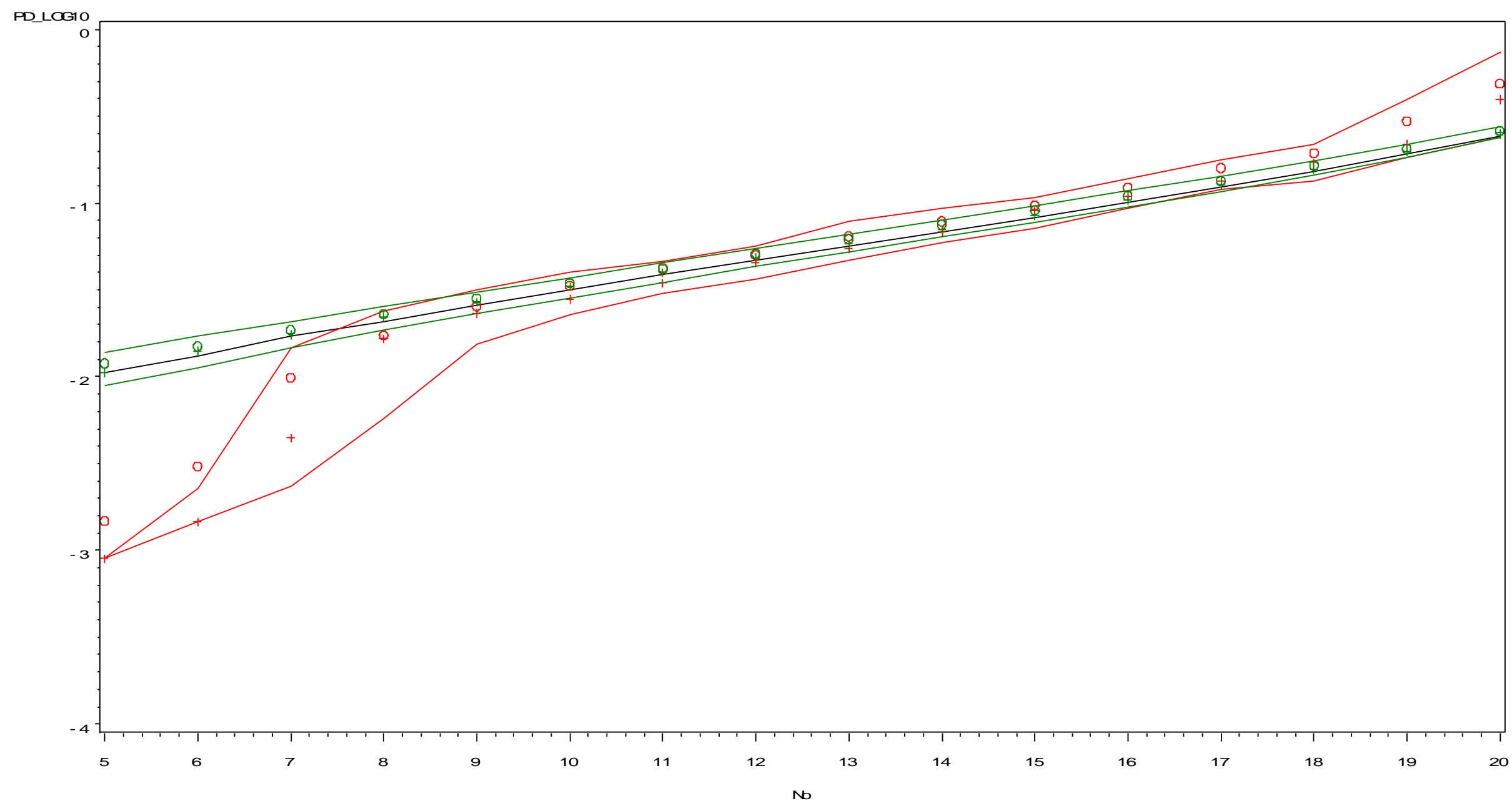

*Performance of the structural vs empirical rating transition matrices in terms of the prediction of the 10-year cumulative PD. X-axis: the number of the initial rating class, Y-axis: 10-year PD (log10-scaled). Black line: "true" (simulated) PD. Red 'o', '+' and '-' lines: mean, median and $25^{th}$/$75^{th}$ percentiles of the empirical PD prediction. Green 'o', '+' and '-' lines: mean, median and $25^{th}$/ $75^{th}$ percentiles of the structural PD prediction.*

In particular, for the middle initial rating classes (10-18), the two approaches turn out to be unbiased in the sense that they both produce, on average, PDs which are very close to the "true" 10-year PD. However, the structural predictors prove here to be more efficient (less volatile), as measured by the distance between the $75^{th}$ and $25^{th}$ percentiles of the predictors across the 100 samples. Thus, even for middle rating classes, the structural predictors stay close to the "true" PDs with a higher probability.

For the more extreme initial ratings (5-10, 19-20), the naïve empirical estimators demonstrated a significantly rising volatility along with considerable prediction biases. The structural estimators remained here unbiased and efficient.

---

[5] For (initial) ratings not present in a sample drawn, the diagonal elements were assumed to equal 100% minus PD, and other elements (apart from default probability) were set to 0.

This demonstrates the clear superiority of the structurally regularized predictors when compared to the naïvely calculated empirical predictors, in small portfolios. The advantage seems to be particularly evident for the better ratings. Intuitively, the multi-year PDs of these ratings depend critically on the probabilities of transitions to the worse rating classes, and thus on the quality of the rating transition matrix.

# Model Extensions

The model and the estimation procedure presented can be extended. Two possible extensions are described below. The first extension deals with being more precise with the assigned PD of an initial rating class in rating transitions, which might be particularly advantageous if the rating classes cover wide PD intervals. The second extension modifies the process specification to apply a link function used in many underlying reduced-type PD rating models (e.g. logit), which also increases the number of the process parameters by one and might therefore offer more flexibility for certain data.

## Increased Precision for Initial PDs in Rating Transitions

This paper proposed in (29) the usage of the assigned rating PD $pd_R^{asn}$ or the mid rating PD $pd_R^{mid} = \frac{1}{2}\left(pd_R^{low} + pd_R^{high}\right)$ as the PD $\widehat{PD}_{i,t}$ of an initial rating $R$ in case of transitions between ratings. The likelihood function in (20) can, however, be highly non-linear with respect to the initial PD $\widehat{PD}_{i,t}$, which deteriorates the approximation.

The approximation can be improved by some assumption of the distribution of $\widehat{PD}_{i,t}$ over the interval $(pd_R^{low}, pd_R^{high})$. This interval may be subdivided into $K$ subintervals $(pd_R^{k,low}, pd_R^{k,high})$ with $k = 1, \dots, K$. Then, the $K$ different transition probabilities/likelihoods using the initial PD $\widehat{PD}_{i,t} = \left(pd_R^{k,low} + pd_R^{k,high}\right)/2$ can be calculated according to (20). These transition probabilities must then be weighted with the assumed probability weights of the subintervals, to arrive at the single transition probability. The simplest assumption would be to use equal weights, which corresponds to the uniform distribution of $\widehat{PD}_{i,t}$ over $(pd_R^{low}, pd_R^{high})$. Another plausible assumption would be to use the unconditional PD distribution derived from the process (3) itself. For the latter, it should be noted that a distant $AP_t$ has roughly normal distribution, resulting from a sum of many (weighted) random returns (see (31) for $Y \to \infty$). More specifically, the unconditional distribution of $AP_t$ is then roughly normal with $E\left(AP_t\right) = \frac{a_0}{1-a_1}$ and $Var\left(AP_t\right) = Var(r_t)/(1 - a_1)^2$. From this, the distribution of corresponding PDs can be determined using (5).

## Alternative Model Specification with a Link Function

Probabilistic reduced-form (non-structural) rating-models (logit/probit) are built on the basis of transformation of a score (normally, a linear combination of predictors such as financial ratios) to a PD via a link function. The score measures the solvency of an obligor and roughly corresponds to the ability-to-pay; the link function is normally some cdf. The relationship is then:

$$PD_t = G\left(-SCORE_t\right)$$

or

$$SCORE_t = -G^{-1}(PD_t)$$

with $G$ as the link function.

For the score, we can then assume a similar autoregressive process:

$$SCORE_{t+1} = a_0 + a_1 SCORE_t + r_{t+1}$$

$$r_t \sim F$$

Then:

$$PD_{t+1} = G\left(-a_0 - a_1 SCORE_t - r_{t+1}\right) = G\left(-a_0 + a_1 G^{-1}(PD_t) - r_{t+1}\right)$$

And the cdf is:

$$P\,(PD_{t+1} < X) = P\left(-a_0 + a_1 G^{-1}(PD_t) - r_{t+1} < G^{-1}(X)\right) =$$

$$P\left(r_{t+1} > a_1 G^{-1}(PD_t) - G^{-1}(X) - a_0\right) =$$

$$1 - F\left(a_1 G^{-1}(PD_t) - G^{-1}(X) - a_0\right)$$

For symmetric return distributions:

$$P\,(PD_{t+1} < X) = F\left(a_0 + G^{-1}(X) - a_1 G^{-1}(PD_t)\right)$$

A similarity to (9) and (10) is obvious. However, with this specification, the model is not invariant as to the scaling of the return. In particular, for a modified specification with a scale factor $S$:

$$SCORE_{t+1} = a_0 + a_1 SCORE_t + S\, r_{t+1}$$

the resulting cdf changes to:

$$P\,(PD_{t+1} < X) = F\left(\frac{1}{S}\left[a_0 + G^{-1}(X) - a_1 G^{-1}(PD_t)\right]\right)$$

The scale parameter $S$ thus offers an additional degree of freedom with this model specification, and the scale can also be estimated from data via ordinary MLE. The distributions $F$ and $G$ can be the same or different.

# Summary and Conclusions

In this paper, we have presented a structural approach linking one-period PD/rating transition probabilities to a theoretically motivated underlying ability-to-pay process. The stochastic process used relies on realistic assumptions of autoregressiveness, mean-reverting, and fat-tails in the return distribution.

The process needs only three parameters to be fully specified. This significantly reduces the statistical degrees of freedom, which proves to be beneficial when dealing with scarce data. The parameters can, in particular, be estimated from observed transitions of (continuous or discretized/rating) PDs using as few as only 50 transition observations. Using the estimated process parameters, along with rating master scale PDs, the structural/regularized one-year transition matrix can be easily obtained and used for various risk management/controlling purposes.

By virtue of the process design, the resulting regularized transition matrices always show the desired properties, such as monotonically decaying (horizontally and vertically) non-zero transition probabilities with a single maximum (∩-shaped), as well as non-intersecting forward and cumulative PD patterns predicted for the initial rating classes. The rating transition probabilities produced are realistic, and generally comply well with the transition patterns observed in typical real-life portfolios.

For small portfolios, with scarce transition data, the approach has been shown to significantly outperform the naïve approach of empirical transition matrices in terms of prediction of multi-year credit metrics. This can be of particular advantage in the context of the newly adopted IFRS 9 accounting standard. Here, the multi-year PDs calculated from the regularized one-year rating transition matrices can be used to precisely calculate the lifetime expected credit losses. Furthermore, the approach allows for an easy embedding of macroeconomic adjustments which are also required by the IFRS 9 standard.

# Appendix: Derivation of pdf and cdf Functions of Future PDs

**Derivation of the probability density function $pdf(PD_{t+1})$:**

For a function $y = g(x)$ the density $f(y)$ of $y$ can be deduced from the density $f(x)$ of $x$ as follows:

$$f(y) = \left|\frac{d\left(\mathrm{g}^{-1}(y)\right)}{dy}\right| f\left(\mathrm{g}^{-1}(y)\right)$$

With following substitutions:

$$y = PD_{t+1} = g(x) = g\left(r_{t+1}\right)$$

$$g(x) = F\left(-a_0 - a_1\left[-\mathrm{F}^{-1}\left(PD_t\right) + x\right]\right)$$

$$r_{t+1} = x = \mathrm{g}^{-1}(y) = \frac{-1}{a_1}\left[\mathrm{F}^{-1}(y) + a_0\right] + \mathrm{F}^{-1}\left(PD_t\right)$$

we have:

$$f(y) = \frac{+1}{a_1}\frac{d\left(\mathrm{F}^{-1}(y)\right)}{dy} f\left(\frac{-1}{a_1}\left[\mathrm{F}^{-1}(y) + a_0\right] + \mathrm{F}^{-1}\left(PD_t\right)\right)$$

Then, taking into account that:

$$\frac{d\left(\mathrm{F}^{-1}(y)\right)}{dy} = \frac{1}{dF\left(\mathrm{F}^{-1}(y)\right)/dy} = \frac{1}{f\left(\mathrm{F}^{-1}(y)\right)}$$

this finally results in:

$$pdf\left(PD_{t+1}\right) = f\left(PD_{t+1}\right) = \frac{1}{a_1}\frac{f\left(\mathrm{F}^{-1}\left(PD_t\right) - \frac{\mathrm{F}^{-1}\left(PD_{t+1}\right) + a_0}{a_1}\right)}{f\left(\mathrm{F}^{-1}\left(PD_{t+1}\right)\right)}$$

**Derivation of cumulative distribution function $cdf(\widehat{PD}_{t+1})$:**

We have:

$$PD_{t+1} = F\left(S - a_0 - a_1\left[-\mathrm{F}^{-1}\left(PD_t\right) + S + r_{t+1}\right]\right)$$

Then:

$$P\left(PD_{t+1} < X\right) =$$

$$P\left(F\left(-a_0 - a_1\left(-\mathrm{F}^{-1}\left(PD_t\right) + r_{t+1}\right)\right) < X\right) =$$

$$P\left(-a_0 - a_1\left(-\mathrm{F}^{-1}\left(PD_t\right) + r_{t+1}\right) < \mathrm{F}^{-1}(X)\right)$$

Assuming that $a_1 > 0$ (positive autocorrelation), this results in:

$$cdf\left(PD_{t+1}\right) = P\left(PD_{t+1} < X\right) =$$

$$= P\left(r_{t+1} > \frac{\mathrm{F}^{-1}(X) + a_0}{-a_1} + \mathrm{F}^{-1}\left(PD_t\right)\right) = 1 - P\left(r_{t+1} < \mathrm{F}^{-1}\left(PD_t\right) - \frac{\mathrm{F}^{-1}(X) + a_0}{a_1}\right)$$

$$= 1 - F\left(\mathrm{F}^{-1}\left(PD_t\right) - \frac{\mathrm{F}^{-1}(X) + a_0}{a_1}\right)$$